\documentclass{ws-mpla}
\usepackage{graphicx}
\newcommand{\Psfig}[2]{\includegraphics[width=#1]{Figs/#2}}
\newcommand{\chibar}{\bar{\chi}}
\newcommand{\sigbar}{\bar{\sigma}}
\newcommand{\FEFF}{{\cal F}_{{\rm eff}}}
\newcommand{\Veff}{V_{{\rm eff}}}
\newcommand{\arcsinh}{\mathrm{arcsinh}\,}

\begin{document}

\markboth{A. Ohnishi, N. Kawamoto, K. Miura}
{Brown-Rho Scaling in the Strong Coupling Lattice QCD}

\catchline{}{}{}{}{}

\title{
Brown-Rho Scaling in the Strong Coupling Lattice QCD
}

\author{\footnotesize A. Ohnishi, N. Kawamoto, K. Miura}
\address{Department of Physics, Faculty of Science, Hokkaido University\\
Sapporo 060-0810, Japan\\
ohnishi@nucl.sci.hokudai.ac.jp}

\maketitle

\pub{Received (Day Month Year)}{Revised (Day Month Year)}

\begin{abstract}
We examine the Brown-Rho scaling for meson masses
in the strong coupling limit of lattice QCD
with one species of staggered fermion.
Analytical expression of meson masses is derived
at finite temperature and chemical potential.
We find that meson masses are approximately proportional
to the equilibrium value of the chiral condensate,
which evolves as a function of temperature and chemical potential.
\end{abstract}

\ccode{PACS Nos.: 
12.38.Gc, 
12.40.Yx, 
11.15.Me, 
11.10.Wx  
}

\section{Introduction}
The origin of masses has been one of the major driving forces in physics.
For hadrons, a large part of their masses are generated
by the chiral condensate.
Since the chiral condensate may vary significantly in hot and/or dense matter,
hadron masses would be also modified.
Brown and Rho conjectured a scaling behavior of hadron masses
in dense medium,\cite{Brown_Rho}
\begin{equation}
m^*_\sigma/m_\sigma \approx
m^*_N/m_N\approx
m^*_\rho/m_\rho\approx
m^*_\omega/m_\omega\approx
f^*_\pi/f_\pi
\ .
\end{equation}
This scaling law (referred to as the {\em Brown-Rho scaling})
suggests that the partial restoration of the chiral symmetry
can be experimentally accessible by measuring in-medium hadron masses,
and triggered many later theoretical and experimental works.
Theoretically, 
a similar behavior is also found in the NJL model\cite{Hatsuda_Kunihiro}
and in the QCD sum rule.\cite{Hatsuda_Lee}
Experimentally, enhancement of dileptons is observed
below the $\rho$ and $\omega$ meson masses
in heavy-ion collisions at SPS\cite{CERES} and RHIC\cite{PHENIX}
and in $pA$ reactions.\cite{E325}
The interpretation of these enhancements is still under debate,
then it is important to examine the Brown-Rho scaling in QCD.
In the lattice Monte-Carlo (MC) simulations,
it is possible to measure hadron masses quantitatively in vacuum
and at finite temperature,
but it is not easy to perform MC simulations at high densities
because of the sign problem.
Furthermore, MC simulations with small quark masses are expensive.

In this work, we investigate meson masses in the strong coupling limit
of lattice QCD (SCL-LQCD) at finite temperature and
density.\cite{Lattice2007-Miura}
In SCL-LQCD combined with the mean field approximation, 
it is possible to obtain analytical expressions of the effective potential
at finite temperature ($T$) and quark chemical potential 
($\mu$).\cite{SCL,Faldt1986}
Hadron masses are studied in SCL-LQCD
at zero temperature,\cite{KS1984,Kluberg-Stern1982}
but not at finite temperatures.
We find that the Brown-Rho scaling for meson masses
approximately holds at finite $T$ and $\mu$ in SCL-LQCD
with one species of staggered Fermion
in the leading order of the $1/d$ expansion.

\section{Meson masses in the strong coupling limit lattice QCD}

In SCL ($g\to\infty$),
we can ignore pure gluonic action proportional to $1/g^2$,
and obtain the following partition function
after integrating spatial link variables
in the leading order of the $1/d$ expansion,\cite{SCL,Faldt1986}
\begin{eqnarray}
{\cal Z} &=& \int {\cal D}[\chi,\chibar,U_0]\,
e^{\frac12 \sum_{x,y}M(x)V_M(x,y)M(y)-S_F^{(t)}-m_0\sum_x \chibar(x)\chi(x)
}
\ ,\\
S_F^{(t)}&=&\frac12\sum_{\mathbf{x},n} \left[
	 e^\mu\chibar(\mathbf{x},n)U_0(x)\chi(\mathbf{x},n+1)
	-e^{-\mu}\chibar(\mathbf{x},n+1)U_0^\dagger(x)\chi(\mathbf{x},n)
	\right]
\nonumber\\
	&\equiv& \frac12
	\sum_{\mathbf{x},n,m}\chibar_a(\mathbf{x},n)
		V^{(t)}_{na,mb}(\mathbf{x})
		\chi^b(\mathbf{x},m)
\ ,
\end{eqnarray}
where the mesonic composite and their propagators are defined as
$M(x)=\chibar_a(x)\chi^a(x)$
and $V_M(x,y)=\sum_{j=1}^d(\delta_{y,x+\hat{j}}+\delta_{y,x-\hat{j}})/4N_c$,
and $d$ denotes the spatial dimension.
We introduce a mesonic auxiliary field ($\sigma$)
through the Hubbard-Stratonovich transformation,
then the action is separated into terms containing
quarks and time-like link variables on the same spatial points,
and it becomes possible to perform the integral
over quark and time-like link variables.
The effective action for $\sigma$ is obtained as,
\begin{eqnarray}
S[\sigma] &=& \frac12 \sum_{x,y}\sigma(x) V_M^{-1}(x,y) \sigma(y)
		+ \frac{1}{T} \sum_{\mathbf{x}}\Veff(\mathbf{x})
\\
  &=& \frac{L^d}{T}\,{\cal F}_\mathrm{eff}(\sigbar)
  +\frac12 \sum_{x,y}\delta\sigma(x)G_\sigma^{-1}(x,y)\delta\sigma(y)
\ ,
\end{eqnarray}
where $N=1/T$ and $L$ stand for the temporal and spatial lattice sizes.

The equilibrium value $\sigbar$ is determined by 
the effective potential minimum,
$\partial{\cal F}_\mathrm{eff}/\partial\sigbar=0$.
In order to obtain the inverse propagator $G_\sigma^{-1}$,
we need to know the interaction term,
$\Veff(\mathbf{x})$,
as a functional of $\sigma(\mathbf{x},n)$.
F{\"a}ldt and Petersson showed that 
$\Veff$ is obtained as a function of $X_N[\sigma]$,
which is a functional of $\sigma_n=\sigma(\mathbf{x},n)$~\cite{Faldt1986},
\begin{eqnarray}
e^{-\Veff/T}&=&\int {\cal D}U_0\,
\mathrm{Det}^{(NN_c)}\left[
  V^{(t)}_{na,mb}+2(m_0+\sigma_n)\delta_{nm}
\right]
\nonumber\\
&=&\int dU_0 
\, \mathrm{Det}^{(N_c)}\left[
X_N[\sigma]\otimes \mathbf{1}_c + e^{-\mu/T}U_0^\dagger +(-1)^N e^{\mu/T} U_0
\right]
\ ,
\end{eqnarray}
where 
$\mathrm{Det}^{(n)}$ denotes the $n\times n$ determinant,
and the temporal gauge
$U_0=\mathrm{diag}(e^{i\theta_1},\cdots,e^{i\theta_{N_c}})$
is adopted in the second line.
$X_N[\sigma]$ is given as,
\begin{eqnarray}
X_N&=&\left|
\begin{matrix}
I_1      & e^\mu	& 0		&\cdots	& & e^{-\mu} \\
-e^{-\mu}& I_2		& e^\mu		&	& & 0 \\
0        & -e^{-\mu}	& I_3		&	& & 0 \\
\vdots	 &	   	&		&\ddots	& &   \\
	 &	   	&		&	&I_{N-1} & e^\mu	\\
-e^\mu	 & 0		& 0		&\cdots	& -e^{-\mu} & I_N\\
\end{matrix}
\right|
- \left[e^{-\mu/T} + (-1)^N e^{\mu/T} \right]
\ .
\label{Eq:XN}
\end{eqnarray}
where $I_k=2(\sigma_k+m_0)$.
Since $X_N$ is expressed in an explicit determinant form,
its derivatives by $\sigma_n$ are also given in the determinant
of smaller matrices.
In obtaining the meson propagator, 
it is enough to evaluate the $U_0$ integral and determinants in 
equilibrium,\cite{SCL,Faldt1986}
and these are given as follows,
\begin{eqnarray}
X_N&=& e^{E/T}+(-1)^Ne^{-E/T}
\ ,\label{Eq:XN0}\\
\Veff
&=& -T\log\left[2\cosh(N_c\mu/T)+\frac{\sinh[(N_c+1)E/T]}{\sinh[E/T]}\right]
\ ,\label{Eq:RN0}\\
\left.
\frac{\partial^2\Veff}{\partial\sigma_n\partial\sigma_{n+k}}
\right|_{\sigma=\sigbar}
&=&
\left[
\frac{d\Veff}{dX_N}
\frac{\partial^2X_N}{\partial\sigma_n\partial\sigma_{n+k}}
+\frac{d^2\Veff}{dX_N^2}
\frac{1}{N^2}\left[\frac{dX_N^{(0)}}{d\sigbar}\right]^2
\right]_{\sigma=\sigbar}
\ ,\\
\left.
\frac{\partial^2X_N}{\partial\sigma_n\partial\sigma_{n+k}}
\right|_{\sigma=\sigbar}
&=&\frac{2}{\cosh^2{E}}\left[
 \cosh{NE}-e^{i\pi k}\cosh[(N-2k)E]
\right]
\ ,
\end{eqnarray}
where $E=\arcsinh[\sigbar+m_0]$ denotes the one-dimensional quark energy.
By requiring null average fluctuation, $\sum_k \delta\sigma_k=0$,
we ignore those terms independent from $k$ in the derivative of $\Veff$.
The Fourier transform of the inverse propagator,
$\widetilde{G}_\sigma^{-1}\equiv\mathrm{F.T.}(G_\sigma^{-1})$,
is then given as,
\begin{equation}
\widetilde{G}_\sigma^{-1}(\mathbf{k},\omega)
=\frac{2N_c}{\kappa(\mathbf{k})}
-\frac{\partial\Veff(\sigbar,T,\mu)}{\partial\sigbar}
\frac{2\sinh{E(\sigbar)}}{\cos\omega+\cosh{2E(\sigbar)}}
\ ,
\end{equation}
where $\kappa(\mathbf{k})=\sum_j \cos k_j$.

\begin{figure}
\Psfig{6.5cm}{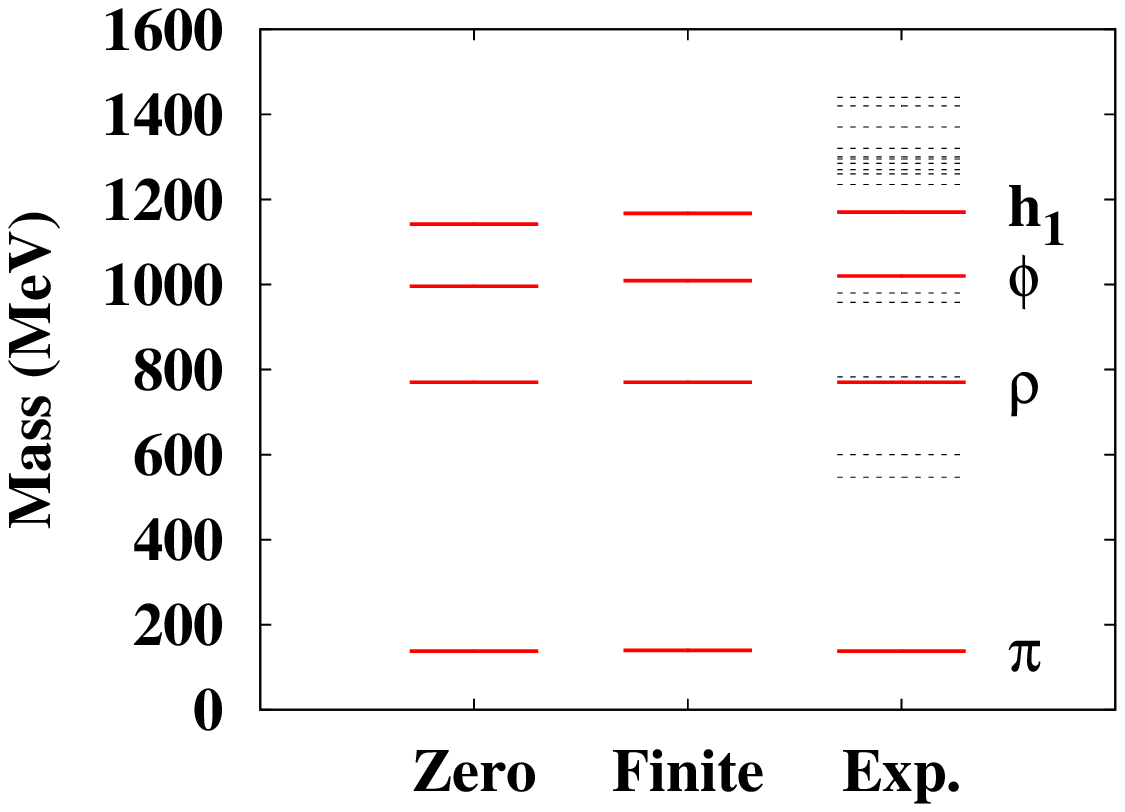}~\Psfig{6.5cm}{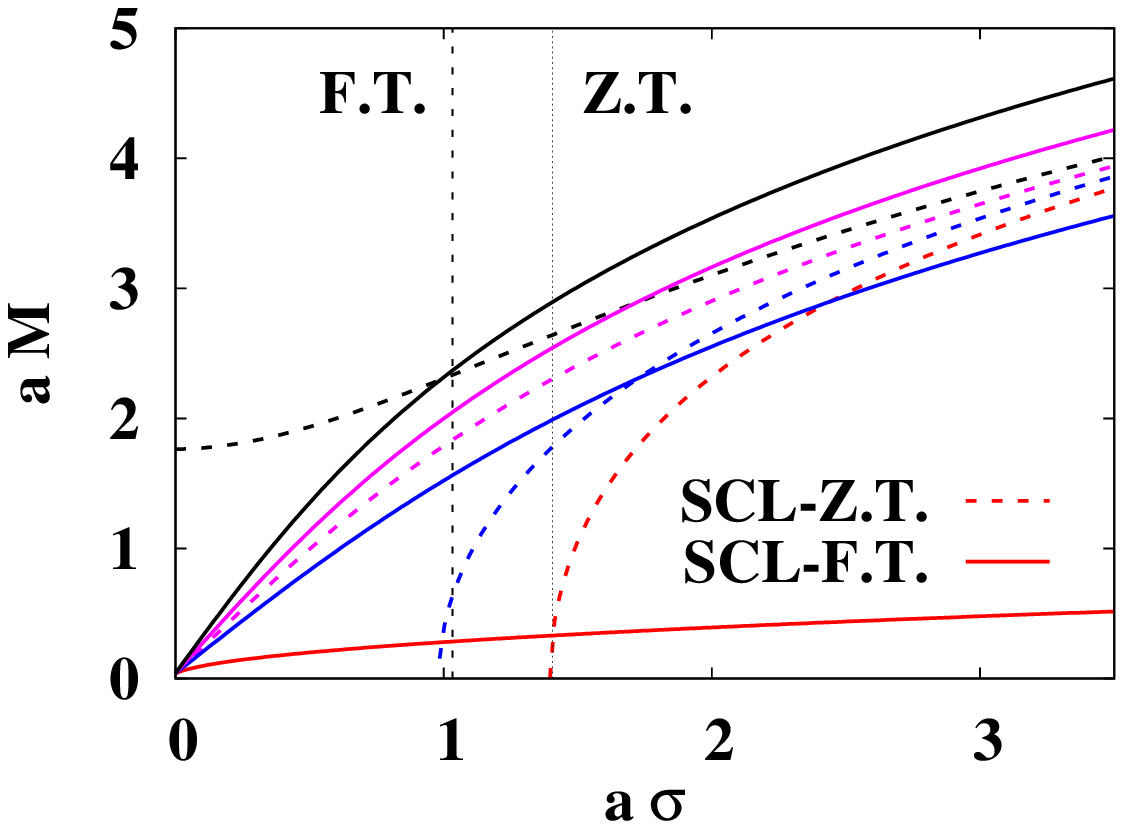}
\caption{
Meson mass spectrum in vacuum (left)
and meson masses as functions of $\sigma$ (right)
in the zero temperature (Z.T.) and finite temperature (F.T.) treatment
in the strong coupling limit of lattice QCD.
}\label{Fig:Vac}
\end{figure}

The meson propagator $\widetilde{G}_\sigma$ depends on $T$ and $\mu$
as well as on $\sigbar$
via the interaction term $\Veff(\sigbar,T,\mu)$.
However, the equilibrium condition $\partial\FEFF/\partial\sigbar=0$
with $\FEFF=N_c\sigbar^2/d+\Veff$
reads $\partial\Veff/\partial\sigbar=-2N_c\sigbar/d$,
and removes the explicit $T$ and $\mu$ dependence, 
\begin{equation}
\widetilde{G}_\sigma^{-1}(\mathbf{k},\omega)
=\frac{2N_c}{\kappa(\mathbf{k})}
+\frac{2N_c\sigbar}{d}
\frac{2\sinh{E(\sigbar)}}{\cos\omega+\cosh{2E(\sigbar)}}
\ .
\end{equation}
As a result, the meson propagator depends on $T$ and $\mu$
only through the equilibrium value of the chiral condensate,
$\sigbar(T,\mu)$.

Meson masses are obtained as the pole energy $\omega=iM + \delta_\pi$
of the propagator at {\em zero} momentum, $\mathrm{k}_j=\delta_\pi$,
where $\delta_\pi=0\ \mathrm{or}\ \pi$.\cite{Kluberg-Stern1982}
This appears from the taste degrees of freedom.
With the choice of $\omega=iM+\pi$, the masses are found to be
\begin{equation}
M=2\,\arcsinh\sqrt{
(\sigbar+m_0)\left(
\frac{\kappa+d}{d}\sigbar+m_0
\right)}
\ ,
\end{equation}
where $\kappa=-d, -d+2, \ldots, d$.

In the chiral limit ($m_0=0$), 
we always have a massless boson for $\kappa=-d$,
as a consequence of the chiral symmetry in the present effective potential. 
For a small current quark mass, we find that
$M(\kappa=-d) \approx 2\sqrt{\sigbar m_0}$,
which may be regarded as the PCAC relation.
Thus we regard the mode $\kappa=-d$ as the pion.
We tentatively assign $\kappa=-1$ corresponds to $\rho$
meson,\cite{Kluberg-Stern1982}
then we can fix the physical scale,
$a^{-1}=497~\mathrm{MeV}$ and $m_0=9.5~\mathrm{MeV}$
by fitting $\pi$ and $\rho$ meson masses in vacuum.
With these parameters and the present assignment,
$M(\kappa=1,3)$ seems to correspond to $\phi$ and $h_1$ mesons
as shown in the left panel of Fig.~\ref{Fig:Vac}.
Contribution from the chiral condensate $\sigbar$
to meson masses (except for $\pi$)
is found to be much larger than that from the current quark mass $m_0$
in vacuum, where $\sigbar_\mathrm{vac} \sim a^{-1}$.

\begin{figure}
\Psfig{6.5cm}{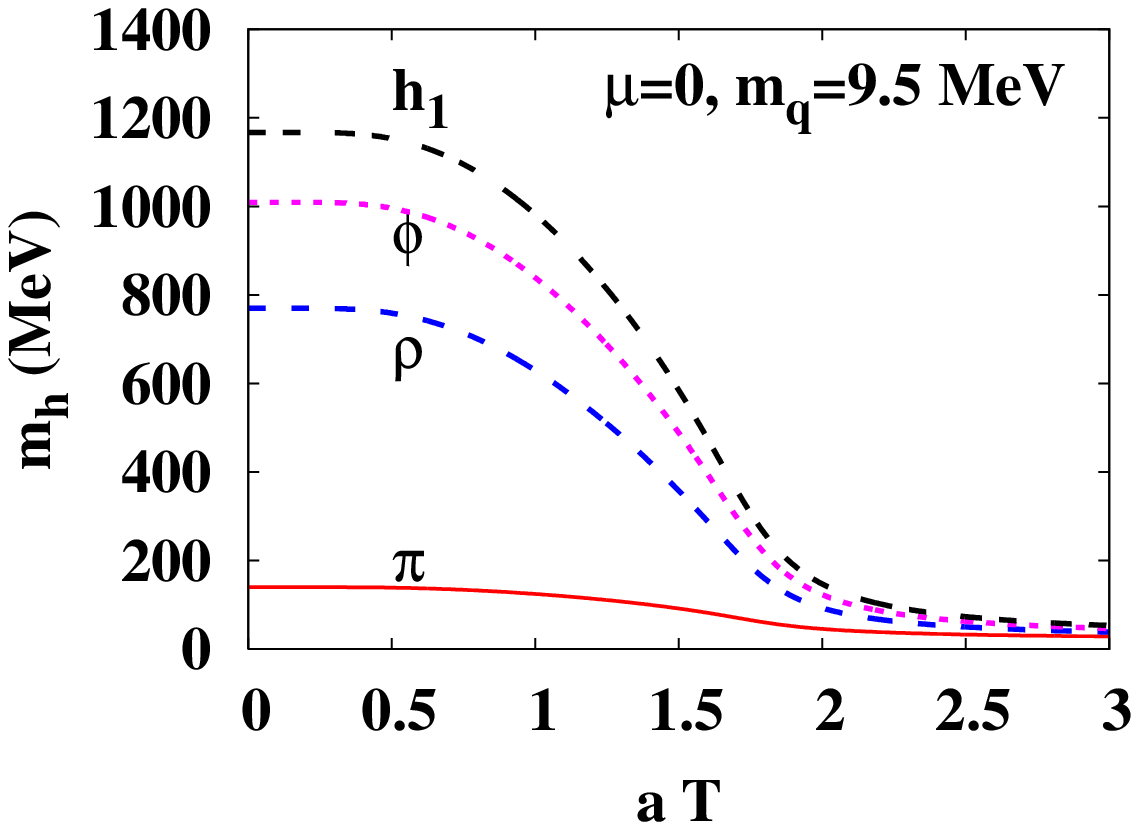}~\Psfig{6.5cm}{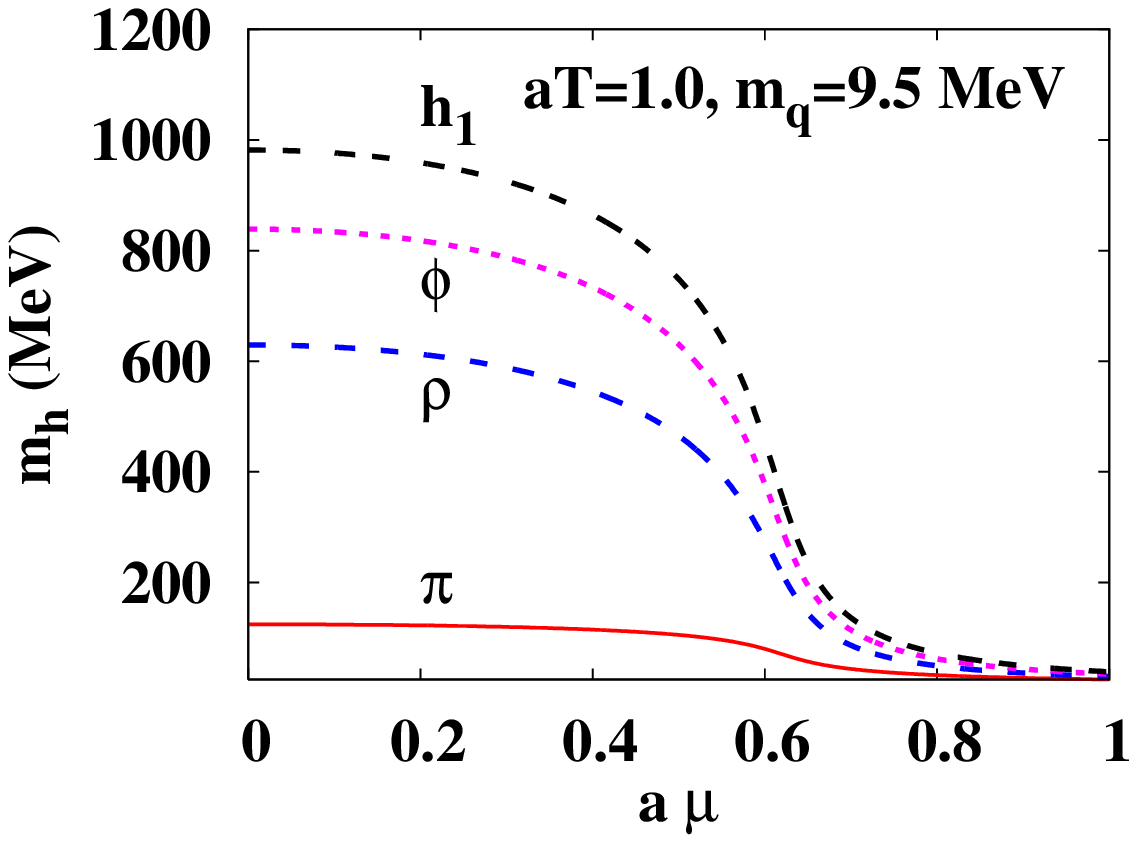}
\caption{
Temperature (left) and chemical potential (right) dependence of meson masses.
}\label{Fig:Tmu}
\end{figure}

In the right panel of Fig.~\ref{Fig:Vac},
we show meson masses as functions of $\sigbar$.
In the range $\sigbar \leq \sigbar_\mathrm{vac}$,
meson masses with $\kappa = -1, 1, 3$ are approximately proportional
to $\sigbar$.
In the present finite temperature (F.T.) treatment,
$\sigbar$ evolves as a function of temperature and density.
As a result, meson masses are also modified
in hot and/or dense matter, as shown in Fig.~\ref{Fig:Tmu}.
These results should be compared
with those in the zero temperature (Z.T.) treatment,\cite{Kluberg-Stern1982}
\begin{equation}
\cosh M = 2(\sigbar+m_0)^2 + \kappa
\ .
\end{equation}
In vacuum, these meson masses explain observed the observed mass spectrum,
but there is no $(T,\mu)$ dependence.

\section{Summary and Discussion}

In this work, we have examined the Brown-Rho scaling for meson masses
in the strong coupling limit of lattice QCD with one species of staggered
Fermion at finite temperature and chemical potential.
Meson masses except for $\pi$ are found to be approximately proportional
to the equilibrium value of the chiral condensate, $\sigbar$.
Since the condensate mode $(\omega, \mathbf{k}) = (0,\mathbf{0})$
corresponds to the chiral partner of $\pi$,
we may assume that $\sigbar$ is proportional to the pion decay constant
in medium, $f_\pi^*$.
Under this assumption, we may conclude that
the Brown-Rho scaling would hold in the strong coupling limit of QCD.

There are many more things to be clarified
including
the meson assignment as pointed out in the symposium,
pion mass behavior around the chiral transition
which is considered to grow in medium,\cite{Hatsuda_Kunihiro}
meson masses with the choice of $\omega=iM$,
too high $T_c$ in SCL,\cite{Ohnishi2007}
and negative eigen values in $V_M$.\cite{MiuraThesis}

\section*{Acknowledgements}
We would like to thank Prof. A. Nakamura for useful suggestions.
This work is supported in part by the Ministry of Education,
Science, Sports and Culture,
Grant-in-Aid for Scientific Research
under the grant numbers,
    13135201,		
    15540243,		
    1707005,		
and 19540252.		

\end{document}